\documentclass[prl,twocolumn,floatfix,groupedaddress,nofootinbib,showpacs,preprintnumbers,
amsmath,amssymb,amsfonts,superscriptaddress,widetable]{revtex4-1}
\usepackage{bm}
\usepackage{mathrsfs}
\usepackage{amssymb}
\usepackage{amsmath}
\usepackage{graphicx}
\usepackage{array}
\usepackage{epsfig,rotating,amsmath}
\usepackage{multirow}
\usepackage{color}
\usepackage{soul}
\usepackage{ulem}
\begin{document}
\normalem

\title{Constraints on Skyrme Equations of State from Doubly Magic Nuclei, Ab-Initio Calculations of Low-Density
Neutron Matter, and Neutron Stars}

\author{C. Y. Tsang}\email{tsangc@nscl.msu.edu}
\affiliation{Department of Physics and Astronomy and National
Superconducting Cyclotron Laboratory, Michigan State University,
East Lansing, Michigan 48824-1321, USA}

\author{B. A. Brown}\email{brown@nscl.msu.edu}
\affiliation{Department of Physics and Astronomy and National
Superconducting Cyclotron Laboratory, Michigan State University,
East Lansing, Michigan 48824-1321, USA}

\author{F.J. Fattoyev}\email{ffattoyev01@manhattan.edu}
\affiliation{Department of Physics, Manhattan College,
                Riverdale, NY 10471, USA}

\author{W. G. Lynch}\email{lynch@nscl.msu.edu}
\affiliation{Department of Physics and Astronomy and National
Superconducting Cyclotron Laboratory, Michigan State University,
East Lansing, Michigan 48824-1321, USA}

\author{M. B. Tsang}\email{tsang@nscl.msu.edu}
\affiliation{Department of Physics and Astronomy and National
Superconducting Cyclotron Laboratory, Michigan State University,
East Lansing, Michigan 48824-1321, USA}

\begin{abstract}
We use properties of doubly-magic nuclei, ab-initio calculations of
low-density neutron matter, and of neutron stars to constrain the
parameters of the Skyrme energy-density functional. We find all of
these properties can be reproduced within a constrained family of
Skyrme parameters. The maximum mass of a neutron star is found to be
sensitive to the neutron effective mass. A value of
[$ m^{*}_{\rm n}/m](\rho_0) = 0.60-0.65 $ is required to obtain a maximum neutron star
mass of 2.1 solar masses. Using the constrained Skyrme functional with the aforementioned effective mass, the predicted radius for a neutron star of
1.4 solar masses is 12.4(1) km and $\Lambda$ = 423(40).


\end{abstract}

\pacs{21.10.Dr, 21.30.Fe, 21.60.Jz, 21.65.-f}

\maketitle

Understanding the nature of dense neutron-rich matter is a major
thrust of current research in both nuclear physics and astrophysics.
Indeed a well-posed question of \emph{``What is the nature of matter
at extreme temperatures and densities?''} is regarded as a new
scientific opportunity for the next decade~\cite{NSACLRP2015}. To
achieve this goal, many experiments and observations are being
carried out using a wide variety of advanced new facilities, such
as, Facilities for Rare Isotope Beams (FRIB), X-ray satellites and
gravitational wave detectors. In interpreting these experimental and
observational results the equation of state (EOS) of neutron-rich
matter plays a critical role. Some parameters of the EOS that are crucial for neutron star properties are not well constrained by nuclear reaction or structure experiment. In particular, the value of
neutron effective mass remains very uncertain\,\cite{Brown:2013pwa,
Li:2018lpy}. In this paper, we focus on the role of the neutron
effective mass on the high density EOS, and show that it can be
tightly constrained using properties of doubly-magic nuclei,
ab-initio calculations of low-density neutron matter in conjunction
with astrophysical and gravitational wave observations. This in turn
leads to constraining nuclear symmetry energy parameters and neutron
skins of medium-to-heavy nuclei.

We start by mentioning the first direct detection of gravitational
waves from the binary neutron star merger GW170817
\cite{TheLIGOScientific:2017qsa} that has already provided a
fundamental new insight into the nature of dense matter. In
particular, this detection provided critical properties of the
neutron equation of state (EOS) that are encoded in the tidal
deformability (also known as \emph{tidal polarizability}) of the
neutron star, an intrinsic property of the star that describes its
tendency to develop a mass quadrupole, $Q_{ij}$, in response to the
tidal field induced by its companion $E_{ij}$,
\begin{equation}
   Q_{ij}=-\lambda  E_{ij}    \ ,
   \label{QvsE}
\end{equation}
where $\lambda$ is the tidal deformability. By comparing point
theoretical mass waveform with the observed neutron star merger
waveform, the gravitational wave data analysis has also revealed the
dimensionless tidal deformability $\Lambda$ of a neutron
star\,\cite{TheLIGOScientific:2017qsa}:
\begin{equation}
 \Lambda \equiv \frac{\lambda c^{10}}{G^{4} M^{5}} = \frac{2}{3}k_{2}\left(\frac{c^{2}R}{GM}\right)^{5} \ ,
   \label{DimLambda}
\end{equation}
where $k_{2}$ is the second Love number\,\cite{Binnington:2009bb,
Damour:2012yf}. The properties of neutron stars including the tidal
deformability and the second Love number is sensitive to
the EOS in the core and can be computed once the EOS is provided. For
complete discussions, on how to calculate tidal deformability please
refer to \,\cite{Hinderer:2007mb, Hinderer:2009ca, Postnikov:2010yn,
Fattoyev:2013yaa}. In this analysis, we will use the most up-to-date
constraint on the tidal deformability\,\cite{Abbott:2018exr}.

The structure of neutron stars is sensitive to the EOS of cold,
fully catalyzed, neutron-rich matter over a range of densities
spanning several orders of magnitude. For the low-density outer
crust we employ the EOS that follows the seminal work of Baym,
Pethick, and Sutherland\,\cite{Baym:1971pw}. In this region, the
neutron-star matter consists of a Coulomb lattice of neutron-rich
nuclei embedded in degenerate electron gas. As the density
increases, the total chemical potential per nucleon of the system
also increases and eventually exceeds the mass of a neutron. At this
point the optimal nucleus cannot hold any more neutrons and the
neutron drip point is reached which defines the interface between
the outer crust and the inner crust\,\cite{Piekarewicz:2018sgy}. The
inner crust consists of complex and exotic structures, collectively
referred to as \emph{nuclear
pasta}\,\cite{Ravenhall:1983uh,Hashimoto:1984}. Due to the great
number of quasi-degenerate low-energy states nuclear pasta systems
display an interesting yet subtle low-energy dynamics that has been
captured using either semi-classical
simulations\,\cite{Horowitz:2004yf, Horowitz:2004pv,
Horowitz:2005zb, Watanabe:2003xu, Watanabe:2004tr, Watanabe:2009vi,
Schneider:2013dwa, Horowitz:2014xca, Caplan:2014gaa} or
quantum-mechanical mean-field approaches\,\cite{Bulgac:2001,
Magierski:2001ud, Chamel:2004in, Newton:2009zz, Schuetrumpf:2015nza,
Fattoyev:2017zhb}. Despite the undeniable progress in understanding
the nuclear-pasta phase a reliable equation of state for the inner
crust is still missing\,\cite{Piekarewicz:2018sgy}. Nevertheless,
one can resort to a cubic spline to interpolate between the outer
crust and the uniform liquid interior which starts at densities of
about half of the nuclear saturation. The neutron star matter then
undergoes a phase transition into a homogeneous liquid core, where
the Skyrme EDFs are applied. More sophisticated crust calculation
exists where interaction terms in the core region are used in the
crustal calculation \cite{Douchin:2001eos}. 
However while some properties of neutron stars, such as ``crustal
radii'' display strong sensitivity to the EOS of the inner
crust\,\cite{Piekarewicz:2014lba}, it was shown that the
\emph{dimensionless} tidal deformability, in particular, is mostly
sensitive to the EOS of the core almost independent of the detailed
EOS of the inner crust\,\cite{Piekarewicz:2018sgy}. Our approach
should therefore yield a reliable result where we keep the
complexity low. 

For the neutron star core, we use the EOS of cold neutron-rich
matter derived from the nuclear Skyrme Energy Density Functional (EDF)
describing the connection among the energy density $\mathcal{E}$,
the pressure $P$, and the baryon density $\rho$ of the system. In
addition, we assume that neutron star-matter is made of nucleonic
matter complemented with electrons and muons in
beta-equilibrium. The pressure of the system can either be found
directly from the nuclear EDF plus leptonic contributions, or from
the energy density and its first derivative
\begin{equation}
P(\rho) = \rho\frac{\partial \mathcal{E}(\rho)}{\partial \rho} -
\mathcal{E}(\rho) \ .
\end{equation}
Neutron stars satisfy the general relativistic stellar structure
equations, also known as Tolman-Oppenheimer-Volkoff (TOV) equations,
\begin{subequations}
 \begin{align}
  & \frac{dP(r)}{dr} = -G \frac{\Big[{\mathcal E}(r)+P(r)\Big]
      \Big[M(r)+4\pi r^{3}P(r)\Big]}{r^{2}\Big[1-2GM(r)/r\Big]} \;, \\
  & \frac{dM(r)}{dr} = 4\pi r^{2} {\mathcal E}(r) \;,
  \end{align}
 \label{TOV}%
\end{subequations}
where $G$ is the gravitational constant, $r$ is the circumferential radius,
and $M(r)$ is the gravitational mass content. Once an equation of state
$(P\!=\!P({\mathcal E)})$ is supplied the TOV equations may be
solved given boundary conditions in terms of a central pressure
$P(0)\!=\!P_{c}$ and $M(0)\!=\!0$. In particular, the mass $M$ and
the radius $R$ are determined from the following two conditions:
$P(R)\!=\!0$ and $M\!=\!M(R)$. Once the TOV equations have been
solved, and the energy density and pressure profiles are obtained,
then one can integrate the differential equation needed to obtain
the tidal deformability\,\cite{Piekarewicz:2018sgy}. 
This TOV solver has been used successfully to connect neutron star properties 
to nuclear matter parameters in Skyrme interactions \cite{Tsang:2019vxn}.

In this paper, we focus on a particular family of the EOS
model---the Skyrme Energy-Density Functionals---due to its
versatility of being able to fit a myriad of nuclear
observables\,\cite{Dutra:2012mb}. In particular, it can be
parameterized to not only reproduce properties of doubly magic
nuclei but also that of ab-initio calculations, which are sensitive
probes of the EOS in neutron-rich
environments\,\cite{Brown:2013pwa}. We start with the results
obtained in\,\cite{Brown:2013mga, Brown:2013pwa}. In Ref.
\cite{Dutra:2012mb} an extensive study was performed to place
constraints on EDFs based on the
properties of nuclear matter. The standard form of the Skyrme EDFs
and the parameters of the Skyrme functional are given in
\cite{Dutra:2012mb}. Out of 240 Skyrme EDFs, the 16
given in Table VI of \cite{Dutra:2012mb} referred to as the CSkP set
best reproduced a selected set of empirical nuclear matter
properties. Five of these were eliminated since they gave
transitions to spin-ordered matter around densities of $ \rho = 0.25
$ fm$^{-3}$. One of the remainder (LNS) produced unstable  finite
nuclei. The remaining 10 are those given in Table I and labeled with
their name and order in Table VI of \cite{Dutra:2012mb}. To this
list we added the commonly used SLy4 \cite{Chabanat:1997un} and SkM*
\cite{Bartel:1982ed} functionals. These 12 EDFs provide a reasonable
range of values for the symmetric nuclear matter (SNM) effective
mass $[m^{\ast}_0/m](\rho_0)$ = 0.70-1.00
($\rho_0 \approx 0.16$ nucleons/fm$^{3}$).
The lower end of this range is that required by
proton scattering on nuclei \cite{bla}.
The upper end is the enhanced value required for the level density of single-particle
energies near the Fermi surface due to the
coupling with surface vibrations \cite{bla}.
They also provide reasonable values for the nuclear incompressibility
($K_{0}  $ = 212-242 MeV) as compared to values extracted from the
energy of the giant monopole resonances ($ K_{0} $ = 217-230 MeV)
\cite{Cao:2012dt} and heavy ion collisions at probe matter 
to densities ranging up to $4.5\rho_0$ \cite{Danielewicz:2002pu}.

In \cite{Brown:2013mga} these 12 EDFs were refit to a common set of
data for nuclear binding energies, charge radii and single-particle
energies from \cite{AlexBrown:1998zz}. It was shown that the EOS for
neutron matter and symmetry energy were constrained at 0.10
nucleons/fm$^{3}$ (about two-third of the nuclear saturation density
for SNM). The slope of the
neutron EOS at this density was not determined as was first pointed
out in Refs. \cite{Brown:2000pd, Typel:2001lcw}. It was also first
shown in Refs. \cite{Brown:2000pd, Typel:2001lcw} that the slope of
the neutron EOS around a density of 0.10 nucleons/fm$^{3}$ was
highly correlated with the neutron skin $ R_{\rm skin} = R_{\rm n} -
R_{\rm p}$ of heavy nuclei such as $^{208}$Pb, where $R_{\rm n}$ and
$R_{\rm p}$ are the root-mean-square radius for neutrons and
protons, respectively.

\begin{figure}
\includegraphics[scale=0.35]{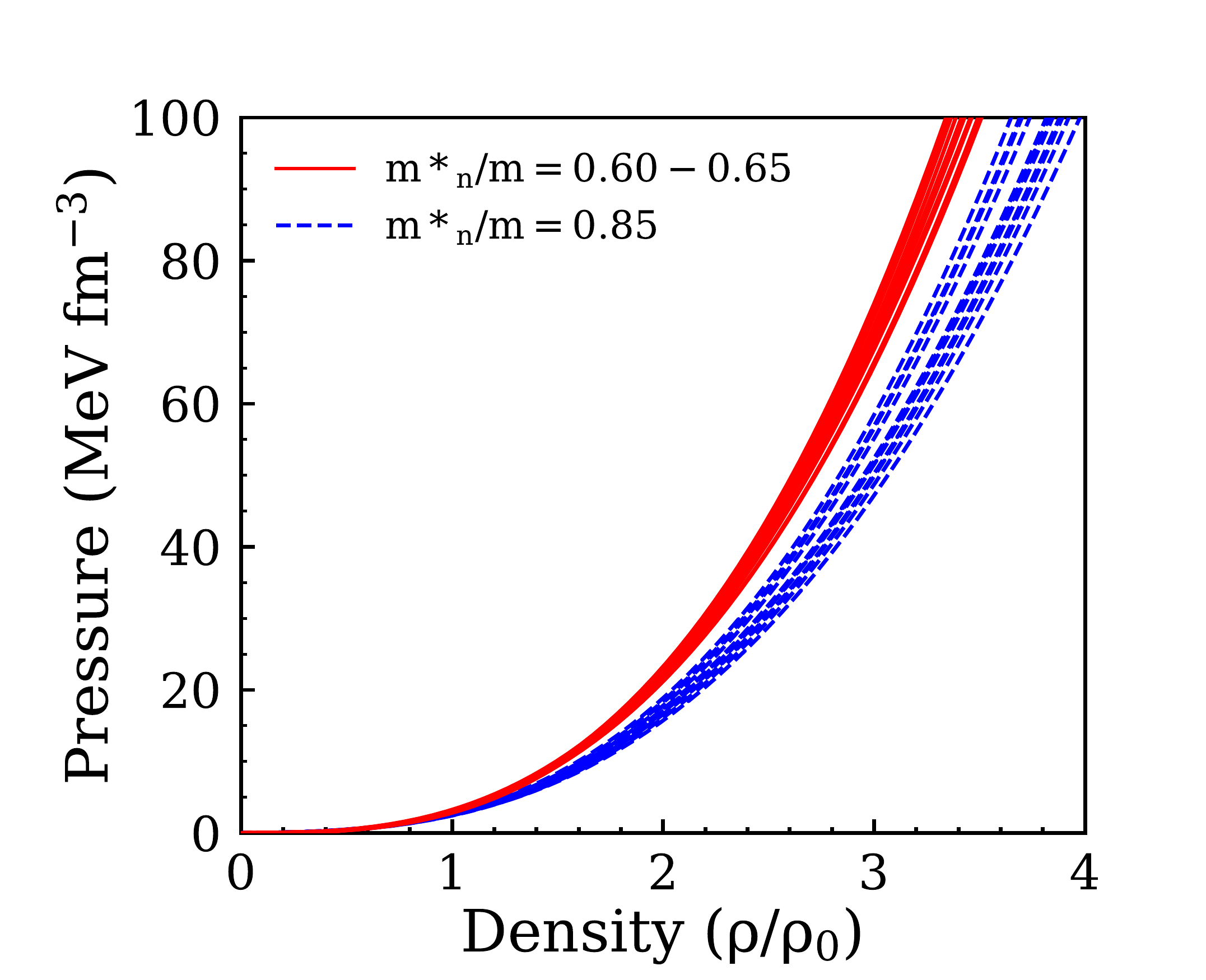}
\caption{The EOS in the form of Pressure versus Density used in this
study. The red band corresponds to 12 Skyrmes with m$^{*}_{n}$/m =
0.60-65 (at $\rho_0$) while the blue band corresponds to 12 Skyrmes with m$^{*}_{n}$/m
= 0.85. All EoSs
are connected to a common crustal EoS \cite{Baym:1971pw}, with
crust-core transition density calculated using empirical relation
of\,\cite{Dutra:2012mb}.} \label{Figure1}
\end{figure}

In \cite{Brown:2013pwa} the same analysis was carried out with the
additional constraint that the neutron EOS reproduced ab-initio
calculations of low-density neutron matter up to the $  E/N  $ of
0.04 neutron/fm$^{3}$
\cite{Hebeler:2009iv,Tews:2012fj, Kruger:2013kua, Gezerlis:2014zia}.
The remarkable result of that
paper was that the parameters of all 12 of the EDFs could easily be
modified to be consistent with both the ab-initio low-density
neutron matter calculations and the large set of nuclear data. The
outcome was that the slope of the EOS could be tightly constrained;
Also, the neutron skins could
be predicted. The largest remaining uncertainty was the neutron
effective mass. In \cite{Brown:2013pwa} a value of
$ [m^{*}_{\rm n}/m](\rho_0)=0.85  $ was chosen, and the blue
dashed curves in Fig. \ref{Figure1}
represent the EOS of this family.

\begin{figure}
\includegraphics[scale=0.35]{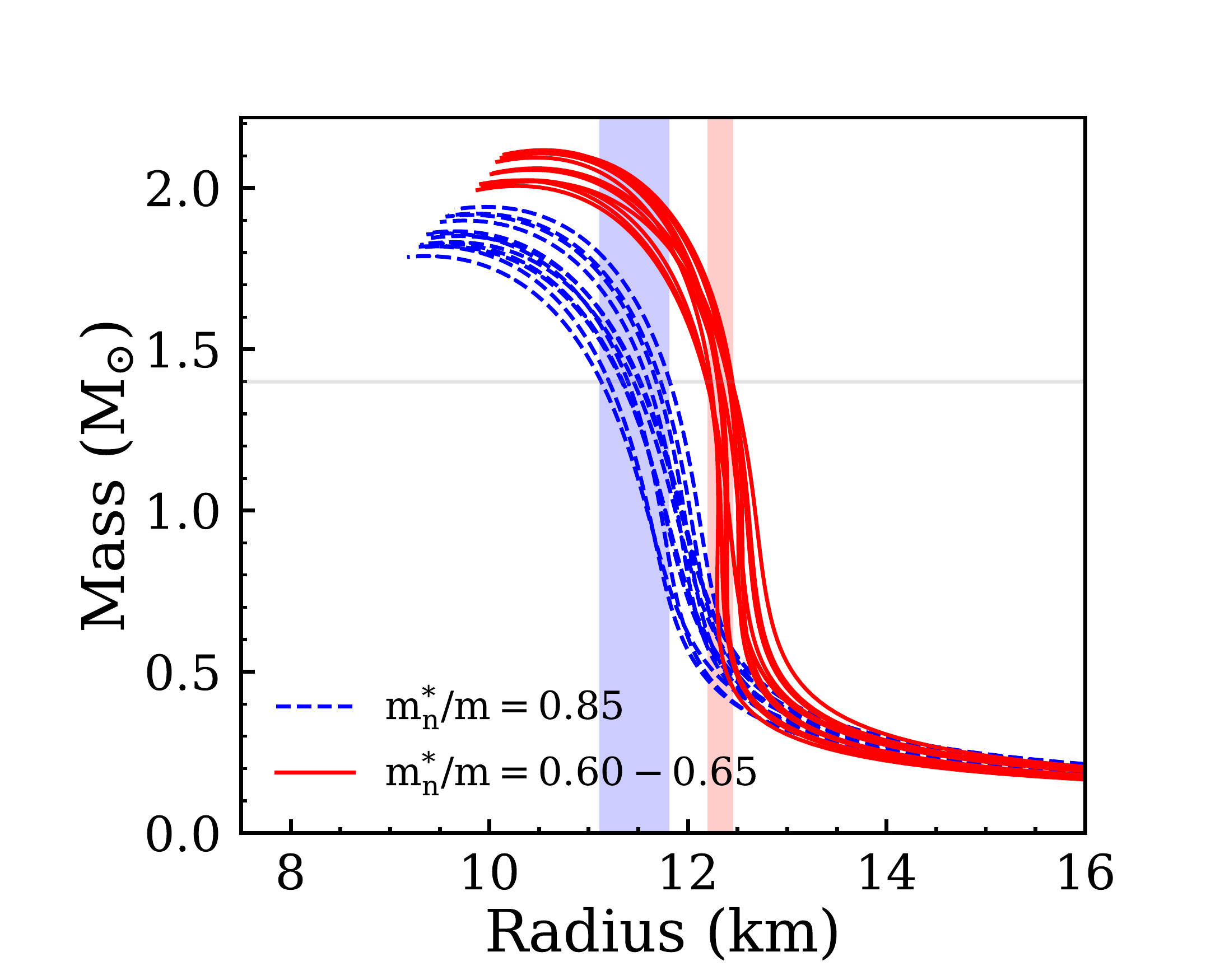}
\caption{ Mass-vs-Radius relation predicted by the two groups of
Skyrme EoSs described in text. The horizontal grey line indicates a value of 1.4 solar mass and the 2 vertical bands show the range of intersections between the grey line and EoSs from each group, which corresponds to the range of predicted 1.4 solar mass neutron star radius.}\label{Figure2}
\end{figure}

\begin{figure}
\includegraphics[scale=0.35]{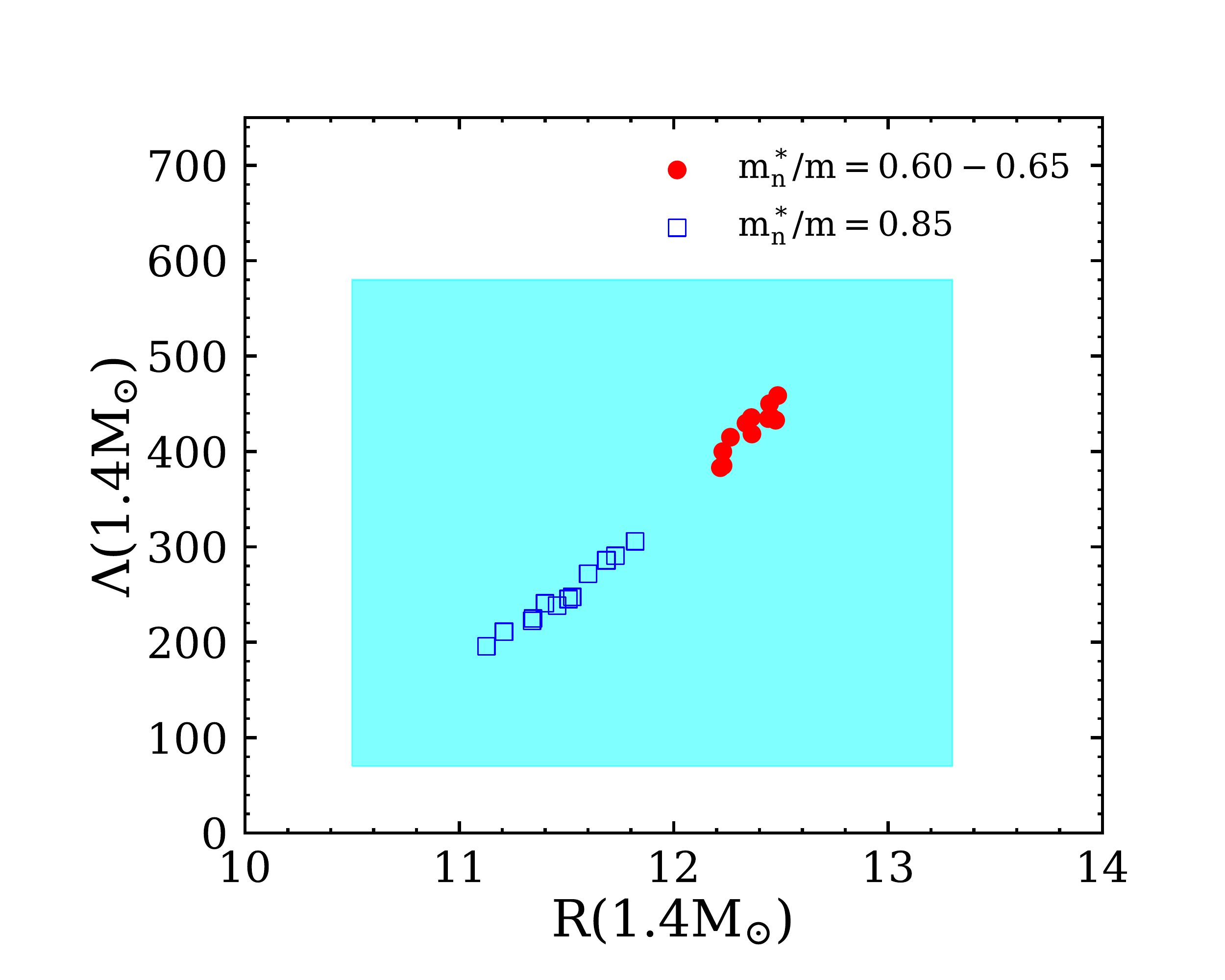}
\caption{ Correlation between neutron-star tidal deformability and
radii of the predicted 1.4 solar mass neutron star from 2 groups of
Skyrmes (blue open square and red solid circle marker). The shaded aqua rectangular
box in the background shows constraints from event
GW170817\,\cite{Abbott:2018exr}. } \label{Figure3}
\end{figure}

We start with this set of 12 Skyrme EDFs from \cite{Brown:2013pwa},
and calculate the properties of neutron stars. The mass-radius
relationship and the deformability-radius relationship for 1.4 solar
mass stars are shown as blue dashed curves in Figs.
\ref{Figure2}-\ref{Figure3}. The predicted values of the tidal
deformability and radii for 1.4 solar mass stars shown as blue solid
symbols are within the constraints obtained from GW170817
represented by a blue shaded square in Figure 3. However, the
maximum mass obtained is 1.8(1) solar mass which is smaller the
2.01(4) solar mass neutron star observed in \cite{Demorest:2010bx,
Antoniadis:2013pzd}. To reconcile the disagreement between our EDFs and this new condition, adjustment to EDFs' parameters is needed. By combining the gravitational and
electromagnetic signals from GW170817 several interesting studies
have been carried out to estimate the maximum mass of neutron stars
that all suggest the absolute maximum mass of a neutron star to be
about $\sim 2.24 M_{\odot}$\,\cite{Ruiz:2017due, Shibata:2017xdx,
Zhou:2017pha, Rezzolla:2018ugo, Zhang:2018bwq, Baym:2019iky}. 

The Skyrme neutron EOS is given by the analytical expression \cite{Brown:2013mga}
\begin{equation}
\mathcal{E}(\rho) =a_{n} \rho^2 + b_{n} \rho^{2+\sigma} + c_{n} \rho^{5/3} + d_{n} \rho^{8/3} \ ,
\label{EnDens}
\end{equation}
where $a_{n}$, $b_{n}$, $c_{n}$,
$d_{n}$ and ${\sigma}$ are constants that depend on the Skyrme parameters. The
first term is from the $s$-wave interaction,
the second term is from the density-dependent $s$-wave interaction,
the third term is the Fermi-gas kinetic energy,
and the fourth term is from the $p$-wave interaction. The
kinetic energy contribution is the $  c_{n}  $ term where $
c_{n}=119  $ MeV fm$^{2}$.

The highest power of the density term  $  d_{n}  $ is related to the
neutron-matter effective mass by
\begin{equation}
\frac{m^{\ast}_{n}(\rho)}{m} = \frac{c_{n}}{c_{n}+d_{n} \rho} \ .
\label{EffMass}
\end{equation}

The next step was to refit the Skyrme parameters to
all of the nuclear data and low-density
neutron EOS constraints considered in \cite{Brown:2013pwa},
with the additional constraint that the maximum neutron star mass
comes out to be about 2.1 solar masses. The outcome is that the
neutron effective mass
at $\rho_0$ is reduced from 0.85 to 0.60-0.65.

This change in the effective mass
has little effect on quality of the fit to nuclear
data or the low-density neutron EOS. For these
12 Skyrme functionals, the rms deviation
for binding energies of $^{40}$Ca, $^{48}$Ca, $^{68}$Ni, $^{88}$Sr,
$^{100}$Sn, $^{132}$Sn and $^{208}$Pb was 0.6 to 0.9 MeV,
and the rms deviation for the root-mean square charge radii of
$^{40}$Ca, $^{48}$Ca, $^{88}$Sr and $^{208}$Pb was 0.015 to 0.024 fm.
Using the constrained Skyrme functional with the aforementioned effective mass,
we obtain $L=65(7)$
MeV for the density derivative of the symmetry energy at a density
of $\rho_0=0.16$ nucleons$/$fm$^{3}$, and neutron skins of $ R_{\rm
skin}(^{208} {\rm Pb}) = 0.194(7)  $ fm and $ R_{\rm skin}(^{48}{\rm
Ca}) = 0.178(3) $ fm.

The results for neutron stars
are shown as red solid curves in Figs. \ref{Figure1}-\ref{Figure2} and red solid circle in Fig. \ref{Figure3}.
As shown in Fig. \ref{Figure1},
the pressure difference between the results with $[m^{\ast}_{\rm n}/m](\rho_0)$=0.60-0.65 and 0.85 for the
neutron effective mass groups are prominent at high density;
 $[m^{\ast}_{\rm n}/m](3\rho_0)$=0.34-0.38 and 0.65, respectively.
It is possible that the effective mass parameter in Skyrme is mocking up
some aspect of dense neutron matter that cannot be extrapolated from 
normal nuclear density EOSs.
In this case the Skyrme phenomenology just provides a convenient and smooth
functional form to be used for the neutron star properties.
It is important to note that all EOSs from both groups satisfy the causality condition.
Their speed of sound never exceeds the speed of light for densities
ranging up to central density of its heaviest permitted neutron star.


\begin{table*}
\begin{center}
\caption{Properties of the fitted Skyrme functionals. The symmetry
energy $  J $, its density derivative $  L  $, the symmetry-energy
incompressibility $  K_{\rm sym}  $, the symmetric-nuclear-matter
incompressibility $  K_{0}  $ and effective mass  $  m^{*}_{0}
$ are evaluated at $  \rho_0=0.16  $ fm$^{-3}$.}
\begin{tabular}{|c|c||c|c|c||c|c|r||c|c|c||c|c|}
\hline name & & $  \sigma  $  & $  K_{0}  $ & $ m^{*}_{0}/m  $ &
$ a_{n}  $
& $  b_{n}  $ & $  d_{n}  $ & $  J  $ & $  L  $ & $  K_{\mathrm{sym}}  $ & $  R_{\mathrm{skin}}  $ & $  R_{\mathrm{skin}}  $ \\
& & &   (MeV) & & (MeV & (MeV & (MeV & (MeV) & (MeV) & (MeV) & (fm) & (fm) \\
& &  & &  & fm$^{3}$) & fm$^{3 \gamma }$) & fm$^{5}$) &  & & &  $^{208}$Pb & $^{48}$Ca \\
\hline \hline KDE0v1 & s3 & 1/6  & 217 & 0.81 &   $  -325  $ &
111 & 472 & 34.6 & 72
& $  -40  $ & 0.200 & 0.178 \\
NRAPR & s6 & 0.14  & 221 & 0.73 & $  -316  $ & 84 & 489 & 34.1
& 70
& $  -46  $ & 0.195 & 0.181 \\
Ska25 & s7 & 0.25   & 220 & 0.98 & $  -281  $ & 37 & 465 &
31.9 &
59 & $  -59  $ & 0.183 & 0.176 \\
Ska35 & s8 & 0.35   & 238 & 0.99 & $  -274  $ & 32 & 467 &
32.0 &
58 & $  -84  $ & 0.184 & 0.177 \\
SKRA & s9 & 0.14  & 213 & 0.80 & $  -347  $ & 143 & 426 &
33.4 &
65 & $  -55  $ & 0.190 & 0.179 \\
SkT1 & s10 & 1/3   & 238 & 0.97 & $  -283  $ & 50 & 476 &
32.6 &
63 & $  -70  $ & 0.190 & 0.179 \\
SkT2 & s11 & 1/3   & 238 & 0.96 & $  -279  $ & 46 & 470 &
32.6 &
62 & $  -75  $ & 0.188 & 0.178 \\
SkT3 & s12 & 1/3   & 236 & 0.97 & $  -275  $ & 32 & 467 &
31.9 &
58 & $  -80  $ & 0.183 & 0.178 \\
SQMC750 & s15 & 1/6  & 223 & 0.75 & $  -307  $ & 76 & 484 &
33.9 &
68 & $  -50  $ & 0.194 & 0.180 \\
SV-sym32 & s16 & 0.30   & 232 & 0.91 & $  -274  $ & 22 & 473
& 31.5 &
58 & $  -77  $ & 0.181 & 0.179 \\
SLy4 & s17 & 1/6  & 222 & 0.76 & $  -299  $ & 68 & 473 & 33.6
&
66 & $  -55  $ & 0.191 & 0.179 \\
SkM* & s18 & 1/6  & 219 & 0.79 & $  -344  $ & 157 & 403 &
33.7 &
65 & $  -65  $ & 0.187 & 0.179 \\
\hline
mean & & & & & &  &  & 33 & 65(7) & -63(24) & 0.194(7) & 0.178(3) \\
\hline
\end{tabular}
\label{Table1}
\end{center}
\end{table*}

When tidal deformability is inferred from our Skyrme EDFs, they show
good agreement with gravitational wave observation as shown
in  Fig. \ref{Figure3}. With the assumed Skyrme functional form
and a neutron effective mass of
0.60-0.65 at $\rho_0$, the $\Lambda$ and radius for a 1.4 solar mass neutron stars
can be narrowed down to 423$^{+35}_{-40}$ and 12.4$^{+0.1}_{-0.1}$
km respectively.

In this paper, we studied the effect of neutron effective mass, the
largest source of EOS uncertainty from nuclear structure, on
neutron star properties.
$[m^{\ast}_{\rm n}/m](\rho_0)$ =
0.60-0.65 is required to produce
a maximum mass of  2.1 solar masses.
We showed that the tidal
deformability $\Lambda$ is sensitive to $m^{\ast}_{\rm n}/m$, and
due to this sensitivity we were able to tighten the constraint on
$\Lambda$ using the Skyrme EDFs that satisfy our effective mass
condition.
This effective mass term, if correct, would strongly affect the
neutron star thermal properties such as its heat
capacity\,\cite{Cumming:2016weq} as well as its neutrino
luminosity\,\cite{Brown:2017gxd, Baldo:2014nem, Reddy:1999eos}. A better knowledge of these
thermal properties would contribute greatly to our understanding of
the neutron star cooling mechanisms\,\cite{Brown:2017gxd}.

More NS mergers
are expected to be detected after the LIGO had resumed its operation, 
and it remains to be seen whether the new constrains
converge to that from these Skyrme EDFs.

We note that the assumed functional form of the neutron EOS from the
Skyrme EDFs provides the analytical connections between its
properties inferred from nuclei (e.g. the value of the symmetry
energy at a density of 0.10 nucleons/fm$^{3}$), those inferred from
ab-initio calculations of low-density EOS of neutron matter, and
those inferred for the high-density pressure of the neutron matter
EOS from the neutron star radii. It will be important to see if any
measured property of nuclei or neutron stars is inconsistent with
our predictions within their error range. If so, then a
less restictive form \cite{bul18} of the EOS will be required.


\begin{acknowledgments}
This work was supported in part by the U.S. National Science
Foundation under Grant Nos. PHY-1565546, PHY-1811855, U.S.
Department of Energy (Office of Science) under Grant Nos.
DE-SC0014530, DE-NA0002923, the Helmholtz Alliance Program of the
Helmholtz Association, contract HA216/EMMI ``Extremes of Density and
Temperature: Cosmic Matter in the Laboratory", and the ERC Grant
No.~307986 STRONGINT. We thank the Institute for Nuclear Theory at
the University of Washington for its hospitality.
FJF is supported in parts by the Summer Grant from the Office of the 
Executive Vice President and Provost of Manhattan College.  
\end{acknowledgments}


%

\end{document}